# A high-efficiency programmable modulator for extreme ultraviolet light with nm feature size based on an electronic phase transition.

Igor Vaskivskyi[1], Anze Mraz[1,2], Rok Venturini[1,3], Gregor Jecl[1,3], Yevhenii Vaskivskyi[1,3], Riccardo Mincigrucci[4], Laura Foglia[4], Dario De Angelis[4], Jacopo-Stefano Pelli-Cresi[4], Ettore Paltanin[4], Danny Fainozzi[4], Filippo Bencivenga[4], Claudio Masciovecchio[4] and Dragan Mihailovic[1,5]

[1]*Jozef Stefan Institute, Dept. of Complex Matter, Jamova 39, SI-1000 Ljubljana, Slovenia, email: igor.vaskivskyi@ijs.si*

[2]*Faculty for Electrical Engineering, University of Ljubljana, Tržaška 25, SI-1000 Ljubljana, Slovenia*

[3]*Faculty for Mathematics and Physics, University of Ljubljana, Jadranska 19, SI-1000 Ljubljana, Slovenia*

[4]*Elettra-Sincrotrone Trieste S.C.p.A., Trieste, Italy*

[5]*CENN Nanocenter, Jamova 39, SI-1000 Ljubljana, Slovenia*

**The absence of efficient light modulators for extreme ultraviolet (EUV) and X-ray photons significantly limits their real-life application, particularly when even slight complexity of the beam patterns is required. Here we report on a novel approach to reversible imprinting of a holographic mask in an electronic Wigner crystal material with a sub-90 nm feature size. The structure is imprinted on a sub-picosecond time-scale using EUV laser pulses and acts as a high-efficiency diffraction grating that deflects EUV or soft X-ray light. The imprinted nanostructure is stable after the removal of the exciting beams at low temperatures but can be easily erased by a single heating beam. Modeling shows that the efficiency of the device can exceed 1%, approaching state-of-the-art etched gratings, but with the benefit of being programmable and tunable over a large range of wavelengths.**

**The observed effect is based on the rapid change of lattice constant upon transition between metastable electronically-ordered phases in a layered transition metal dichalcogenide. The proposed approach is potentially useful for creating tunable light modulators in the EUV and soft X-ray spectral ranges.**

In the last century, the range of useful wavelengths of coherent radiation was significantly extended from infrared, visible and ultraviolet to extreme ultraviolet (EUV), soft- and hard X-rays. While in the optical range it is possible to effectively manipulate the spectral, temporal, or spatial content of the light using various passive (lenses, mirrors, diffraction gratings) and active (acousto- and electro-optic modulators, spatial light modulators – SLM, etc.) elements, this becomes a challenging task for high photon energy beams. Several approaches can be used to produce EUV and X-ray diffraction gratings, such as lithographic etching of thin-film transmission gratings[1] or designing special blazed multilayer gratings[2]. However, the working spectral range of these devices is usually relatively narrow and their manufacturing is a technologically complicated task due to the need for precise control of the etching process.

Temporal modulation of the X-ray beam is usually performed by mechanically moving one or more optical elements in the beam path. Such an approach is cumbersome, has limited speed and precision, and is not always possible due to geometrical constraints. Recently, a miniature microelectromechanical system (MEMS) was presented, showing 350 MHz modulation frequency[3]. However, the proposed device is complicated, its working frequency has to be matched to the repetition rate of the light source and it requires a beam with a small footprint, which might limit the total photon throughput. A new proof-of-principle device based on the phase change material $Ge_2Sb_2Te_5$ was proposed for a reflective SLM in the X-ray spectral range[4]. While showing fast "Writing" speed and high stability, this device suffered from a rather large feature size, limited by the heat propagation effects and the thermal nature of the switching process.

Here we demonstrate a novel approach, which can be used as a reflective SLM for the EUV or soft X-ray beams, featuring ultrafast switching time and sub-90 nm feature size. For the proof-of-principle demonstration, we use two crossed coherent EUV beams ("Write" beams) to imprint a metastable periodic structure into the electronically ordered quantum material $1T-TaS_2$, in which the lifetime of the electronic state is controllable[5] (Figure 1a). This results in a diffraction grating, which efficiently deflects the short wavelength photons.

Transient gratings (TGs), which are formed by the interference of non-collinear pulsed visible or infrared laser beams interacting within a material, are commonly used to investigate carrier diffusion and lifetimes[6–8], heat propagation effects[9,10], acoustic waves[11,12], phonon polaritons[13,14], electron-, vortex dynamics[15], and spin dynamics[16–18] in a wide range of materials. Recently, the EUV TG approach was demonstrated at the Fermi FEL source[19–21]. Such gratings are transient since they are limited by the nonequilibrium lifetime of the intrinsic excitations created in the material by the laser pulses.

Typically heat dissipation, carrier diffusion, or the charge-carrier recombination processes limit the grating lifetime to short timescales, usually not exceeding the ns range. This can be overcome by the use of a metastable electronic state in 1T-TaS$_2$ with a tunable lifetime (from microseconds at 150 K and exceeding typical laboratory scales below 70 K[5]).

This material is a layered Van der Waals system (Fig. 1b) that exhibits a series of electronic ordering transitions and exotic metastable states in response to exposure to light. Under quasi-static thermodynamic conditions, the high-temperature metallic state develops an incommensurate (IC) charge density wave (CDW) at 550 K, which transforms to a unique 'nearly commensurate' (NC) state at 350 K that persists down to 180 K on cooling, whereupon it undergoes a first-order transition to a commensurate (C) CDW. The NC state is best thought of as a regular patchwork of commensurate domains, separated by domain walls, shown schematically in Fig. 1c. Associated with the NC-C transition is a significant change in lattice constants[22], namely the out-of-plane expansion is particularly large and hysteretic, as shown in Fig. 1d.

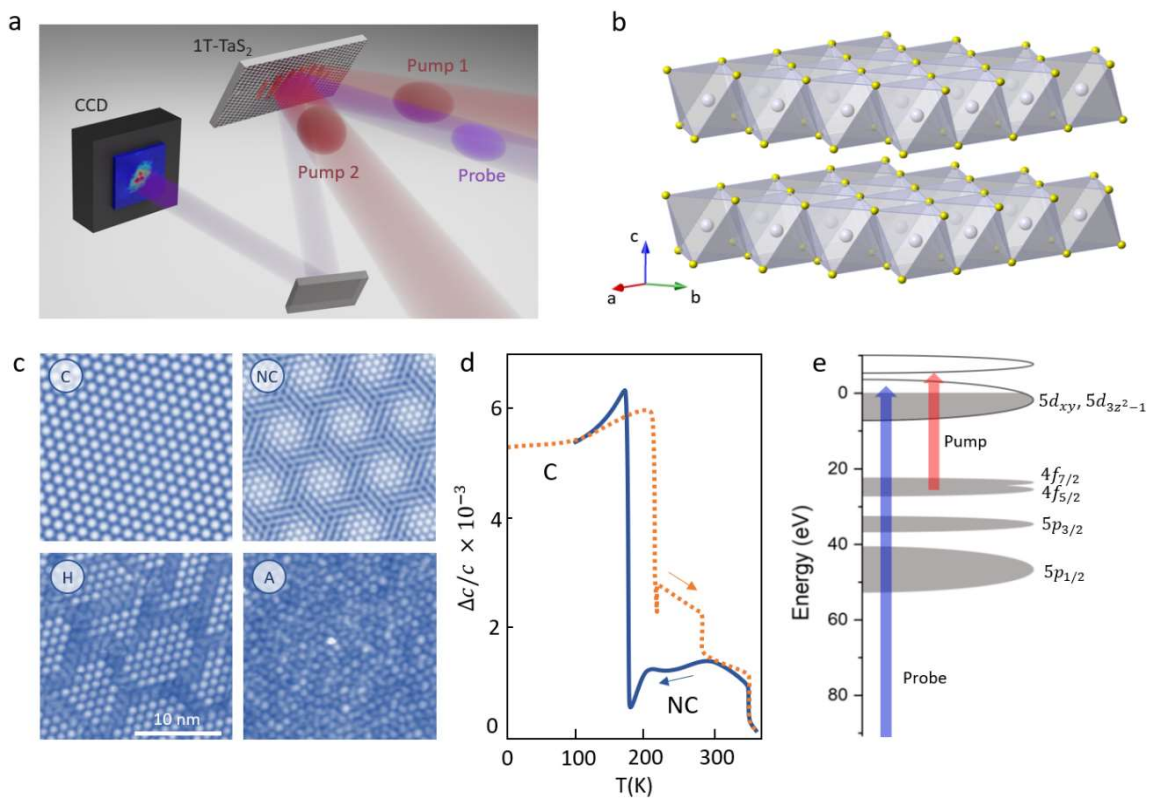

Figure 1 a) The optical layout showing the EUV write beams (Pump 1,2), and the probe beam. b) The crystal structure of 1T-TaS$_2$. c) A schematic representation of the electronic ordering in the thermodynamic C and NC CDW states, and the photoinduced states H and A. d) The lattice expansion along the c axis (after Sezermann et al[22]) for cooling (blue) and heating (orange) cycles. e) An energy level diagram showing the Pump and Probe transitions. The assignments are from Ettema et al[23].

Optical excitation of the C state results in the formation of either a metastable chiral domain (H) state[23] or an amorphous (A) Wigner glass state[24], depending on the excitation parameters[25] (Fig. 1c). The lifetime of the 'hidden' H state, which is superficially similar to a supercooled NC order, can be conveniently controlled by temperature and strain[5]. The A state is more thermally robust than the H state and appears to exist on typical laboratory timescales up to at least 200 K[24].

In this work, we show that the excitation of $1T\text{-}TaS_2$ with two crossed EUV beams at low temperatures creates a metastable diffraction grating that deflects EUV light with remarkable efficiency. We also demonstrate that the imprinted structure can be erased with a single EUV beam, resulting in a programmable modulator for coherent EUV radiation. The extraordinarily high diffraction efficiency of the device is caused by the lattice contraction in an out-of-plane direction associated with the electronic phase transition.

## Results and discussion

We study the device operation in a typical TG experimental configuration, focusing on the evolution of the electronic order in $1T\text{-}TaS_2$ at two temperatures. First, we perform measurements at room temperature where the ground state of the sample is NC CDW and no long-lived metastable state is expected. We observe a typical TG response on the picosecond time scale (Fig. 2a), which is driven by the nearly-resonant pump with the photon energy of 31.1 eV and probed away from any atomic transitions by 93.3 eV photons in backward diffraction geometry (Fig. 1a). The TG period was set to 87 nm. Typically for EUV, the signal consists of an exponential background with superimposed oscillations, which originate from periodic modulation of lattice temperature[26]. This is additionally confirmed by the rise time of the signal (5.2 ± 1 ps), which matches with the electron-lattice thermalization time[27]. The relaxation rate, on the other hand, is given by the heat diffusion time in the system. The slower oscillation corresponds to the surface acoustic wave, with a frequency close to the TA phonon (Fig. 2c), while the faster oscillation might be attributed to the surface skimming longitudinal wave with a frequency close to the LA phonon. After 200 ps the TG signal almost completely disappears, signifying the thermal equilibration between the "bright" and "dark" grating fringes. The disappearance of the diffracted signal, however, does not necessarily mean the complete relaxation of the system to the ground (NC) state. Earlier all-optical studies[27] have revealed that at high excitation fluences the lattice might momentarily reach the NC-IC transition temperature, so it is plausible that the whole probed area of the sample is thermally converted into the high-temperature IC state, which then relaxes back to the NC state on a much longer timescale, once the heat is transported away. It is worth noting that due to non-resonant probe photon energy, we do not directly observe sub-picosecond change of electronic order[28]. Contrary, in all-optical TG experiment, the fast electronic component is about 10 times stronger than the slower lattice response (cyan trace in Figure 2a).

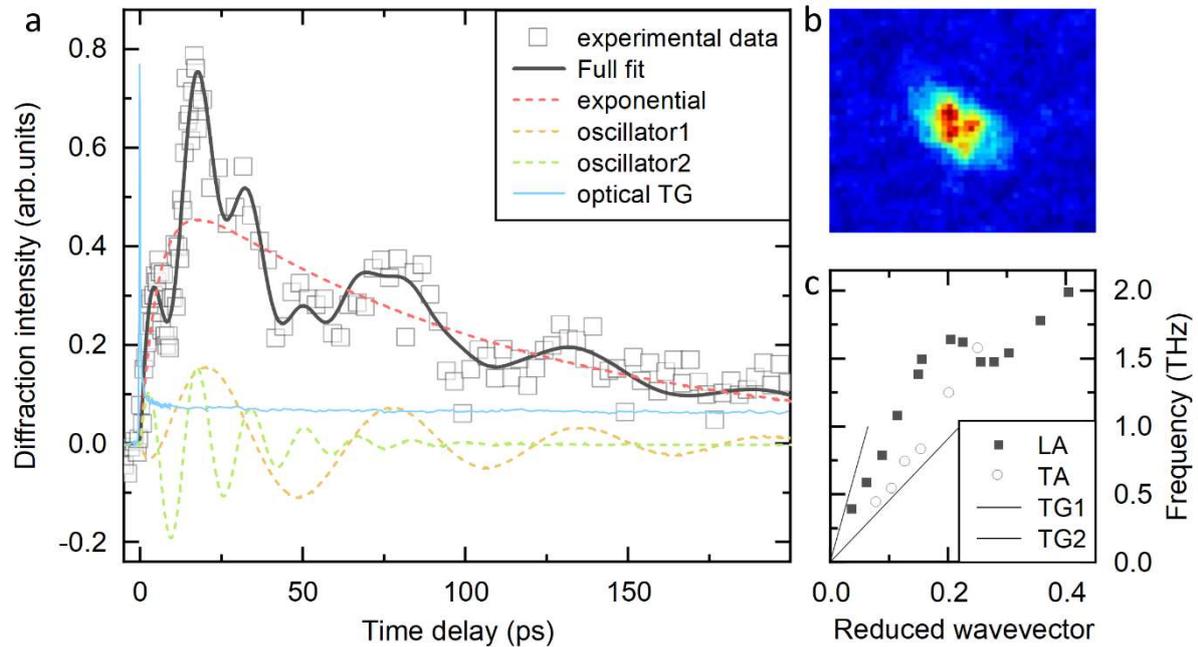

Figure 2 a) The intensity of the probe beam diffracted from the EUV TG created by the Write beams. The fit is shown by the black line. Dashed lines show separate components of the fit. Blue solid line: optical pump-optical probe TG experiment. b) The footprint of the diffracted probe beam on the CCD detector. c) The acoustic phonon velocities deduced from the grating period and oscillation frequency are indicated by the lines. The data points are the LA and TA acoustic phonon frequencies from neutron scattering data[29].

Next, we perform a similar experiment at 100 K, where the ground state of 1T-TaS$_2$ is a C CDW. At this temperature, the pulsed photoexcitation is known to create various metastable orders. Already after the first exposure to the crossed pump beams, a strong scattering appears on the detector at the wavevector of the TG. Remarkably, the scattering intensity was still present and almost constant even after the pump beams are blocked, revealing permanent "imprinting" of the periodic structure (Fig. 3), which does not relax for at least 750 s. After 750 s, the sample was excited by *one* of the EUV pump beams, which does not create an interference pattern in this case, but rather plays a role of a heating pulse. We show in Fig. 3b that this "erases" the diffraction grating in 48±3 s (exponential fit). After blocking the pump beam and allowing the sample to cool down, the whole process of semi-permanent switching can be repeated on the same spot of the sample, confirming that no degradation of the surface occurred.

We attribute this behavior to a photoinduced phase transition to a metastable hidden electronic order [23,24,27] as schematically illustrated in Fig.3a: while the bright illuminated stripes in the EUV grating are converted to the photoinduced state, the dark areas stay in the ground state. The diffraction efficiency of the imprinted structure is at least one order of magnitude higher than the TG at 300 K. Such a stabilized structure would not be possible in the event of a thermal phase transition, in which case the temperature of the "bright" and "dark" stripes would equilibrate on a sub-ns time-scale and thus the whole sample would end up in a single phase, either NC or C CDW depending on the final lattice temperature. Therefore, the existence of a photoinduced hidden state is essential for such a long-lived effect to occur. While from the present data we do not have direct information on the type of the final state, judging from the high stability of the structure at 100 K, it is plausible that the A phase or a mixture of the A and H states is reached.

We note that the EUV light used in the present work is very different from the traditional optical or electrical excitations, therefore it might not be possible to define the type of the photoinduced state based solely on existing values of excitation fluence, and systematic study of the switching efficiencies at different wavelengths is required to elucidate the phenomenon. In our experiment, the fluence in the bright fringes of the crossed pump beams reached 15.2 mJ/cm$^2$, which is significantly higher than the excitation fluences used in earlier optical switching experiments and together with shorter absorption length of resonant EUV light, this corresponds to ~30 times higher energy density absorbed in the sample. On the other hand, the density of absorbed photons is comparable in the present case and earlier optical data.

Similarly to the room-temperature case, the diffraction signal at 100 K mainly originates from the periodic deformation of the lattice. The rather inhomogeneous shape of the diffracted signal observed in such a metastable configuration (Fig. 3c) can be attributed to a non-uniform heating and strain in partially cracked or detached flakes on the sample surface, which affects the stability of the photoinduced state[30] and causes its partial relaxation.

Two pathways for thermal erasing of the semipermanent periodic structure with a single pump beam can be envisaged. (i) If the fluence of the pump beam is above the switching threshold, then each individual pulse causes the whole sample to be converted into the photoinduced state. This would cause immediate disappearance of the diffracted signal. At later time, due to excessive DC heating from the long pulse train, the sample might recover to the ground state. (ii) If the fluence of the single pump beam is below the switching threshold, then the DC heating of the sample is the only necessary process to be considered. The erasing time would depend on the heat load and the resulting sample temperature[5].

Since in the present experiment with 3.8 mJ/cm$^2$ pump the diffraction intensity drops with relatively long time constant (48±3 s), we conclude that the scenario (ii) is operative.

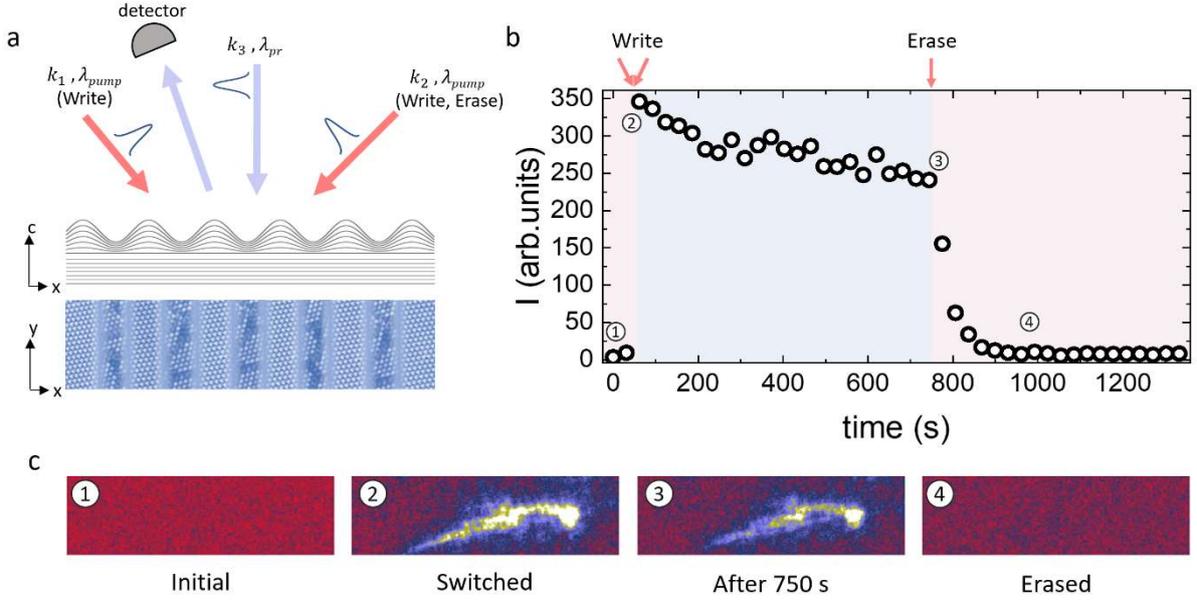

*Figure 3 a) A schematic diagram of the sample dilation along the c axis caused by writing a grating onto the sample and the metastable periodic C/H structure. b) The intensity of the diffracted probe beam on the CCD as a function of time, showing the Write and Erase pulse timing. c) The diffracted streaks recorded on the CCD at the numbered data points in panel (b): 1 – initial state; 2 – immediately after switching; 3 – 750 s after switching; 4 – after erasing the grating with a single EUV beam.*

To understand the unexpectedly high diffraction efficiency of the metastable structure, we have to consider the evolution of the crystalline lattice, which accompanies the electronic phase transition. The C state is characterized by the unique ordering of the CDWs in the neighboring layers, which, among others, causes a ~0.5% change of the out-of-plane lattice constant when cooling from the NC state (Fig. 1d). Since the photoinduced states are characterized by NC-like stacking[31–33], a jump of a similar magnitude can be expected also in the present case resulting in a real-space electronic and lattice modulation. This can be considered as a nearly-sinusoidal reflective diffraction grating (Fig. 3a). Its efficiency can be estimated from[34]:

$$\eta = \frac{I_d}{I_0} = R J_1^2 \left\{\frac{a}{2}\right\},$$

where $I_d$, $I_0$ are diffracted and incident intensities respectively, $R$ is the reflectivity of the material, $a$ is the peak-to-peak excursion of the phase and $J_1$ is Bessel function of the first kind. In paraxial approximation $a = \frac{4\pi h}{\lambda}$, where $h$ is the height of the surface modulation and $\lambda$ is the photon wavelength. The resulting diffraction efficiency for 13.3 nm photons in our experimental geometry is $\eta_{calc} = 10^{-7}$, this matches well the experimentally measured value $\eta_{exp} = 9.5 \times 10^{-8}$. Here, tabulated optical constants[35] were used for calculating the reflectivity of the sample and the thickness of the switched layer. Importantly, according to this simple modeling, by tuning the experimental geometry and impinging the laser beams at an oblique angle, it should be possible to enhance the efficiency of the grating by more than 2 orders of magnitude. The angular dependence of the diffraction efficiency for different photon energies is plotted in Fig. 4a. For shorter values of $\lambda$ and at large incidence angles the efficiency can be enhanced further, and for $\lambda = 0.5$ nm the calculated value reaches $\eta_{calc} = 0.1\%$. It is also worth noting that unlike the polished optical elements, the proposed device can be produced with an atomically-flat surface owing to the layered structure of 1T-TaS$_2$.

We estimate the amplitude of the grating to reach $h = 0.1$ nm in the present experiment, which is significantly larger than previously reported values for EUV TG[36]. This parameter is responsible for the high diffraction efficiency of the structure and is defined by the relative change of the unit cell at the phase transition and the attenuation length of the pump photons in the sample. By switching a thicker layer of the sample, it should be possible to reach $\eta_{calc} > 1\%$ (Fig. 4b) and thus approach the efficiencies of the state-of-the-art multilayer etched gratings (~5%)[1]. In general, it might be beneficial to use shorter wavelength photons due to their longer absorption length. However, further investigation of the switching process at different atomic resonances and with non-resonant excitation is required to optimize the wavelength. By optimizing the experimental conditions and fine-tuning the balance between thermal and non-thermal effects, it might be also possible to change the duty cycle and spatial profile of the imprinted structure, which can help to redistribute the intensity between different diffraction orders. We expect that the feature size of the grating can be further decreased approaching the intrinsic domain size in the hidden state (10-20 nm). The phase transition was already successfully triggered in a 60 nm device by current injection[37].

While in the present experiments we used bulk exfoliated crystals, which is not a technologically scalable solution, recent developments in chemical vapor deposition[38,39] molecular beam epitaxy (MBE)[40] growth of thin 1T-TaS$_2$ might provide the prospects for industrial manufacturing of high-quality and large-area devices based on thin films.

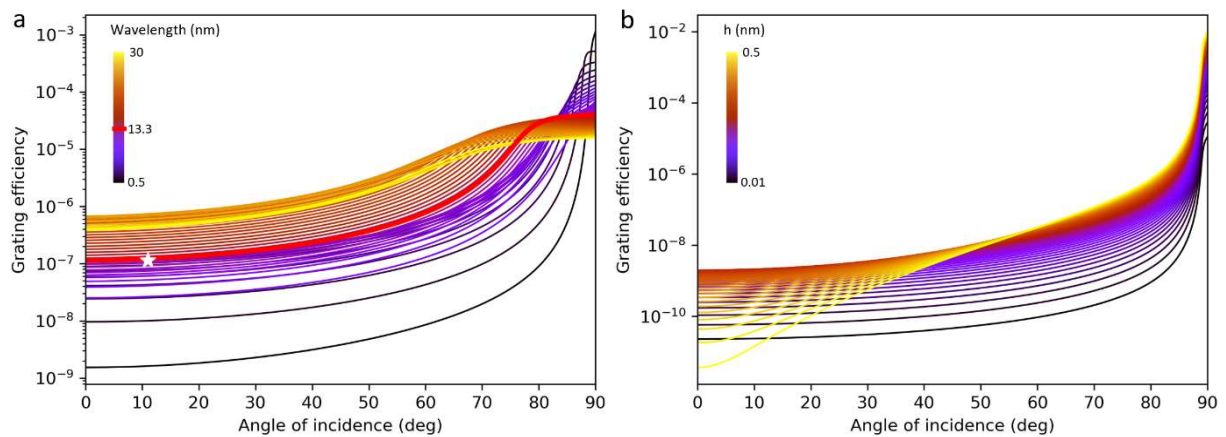

*Figure 4 a) Calculated diffraction efficiency of the sinusoidal grating as a function of the angle of incidence for different wavelengths in the range 0.5-30 nm. Bold red line - the modeling for the probe wavelength λ=13.3 nm used in the experiment. The white star indicates the experimentally measured value. b) Calculated diffraction efficiency of the sinusoidal grating as a function of the angle of incidence for fixed wavelength λ=0.5 nm and different modulation amplitudes in the range 0.01-0.5 nm.*

# Conclusions

We have created a programmable EUV and soft X-ray spatial light modulator in which the periodic structure can be written and erased on demand using short laser pulses. The unexpectedly large amplitude of the surface modulation arises from the large out-of-plane lattice contraction concurrent with the photoinduced phase transition. The stability of the structure is directly related to the long lifetime of the photoinduced hidden charge-ordered states in the material. The formation of the grating is not inherently limited to the present combination of wavelengths. An extension to other wavelengths would result in a wavelength-tunable and programmable grating for EUV and soft X-ray radiation.

Beyond the laser imprinting, which requires two crossed coherent EUV beams for "Writing" the structure, a hybrid electro-optical device could be envisaged. In this case, the textured electronic structure could be created by current injection using an array of electrodes in a manner very similar to liquid-crystal-based SLMs. The presented tunable device can be also used as an intensity and spectral monitor, dispersing a small portion of the beam onto the reference detector. By varying the properties of the imprinted grating, it would be possible to tune the usable spectral range and/or the spectral resolution of a spectrometer without the need to replace or mechanically move any component of the system, adding much more flexibility in EUV and soft X-ray optical layouts.

**Author contributions.** IV, AM, RV, GJ, YV, RC, FP, LF, DA, JPC, EP and DF performed TG experiments; IV, CM and DM devised the experiments; IV, AM, RV, GJ, YV and DM performed the analysis, IV and DM wrote the manuscript; IV led the experimental team; all authors contributed to revising the manuscript.

**Acknowledgments.** We thank for the support from the Slovenian Research and Innovation Agency P1-0040, A.M. to PR-08972, R.V. to PR-10496, G.J. to PR-11213, Slovene Ministry of Science (Raziskovalci-2.1-IJS-952005), E.P. to the European Union's Horizon 2020 research and innovation program under the Marie Sklodowska-Curie Grant Agreement No. 860553. We wish to thank Petra Sutar for providing the samples.

**Data availability.** The data used in this study are available from the corresponding author on reasonable request.

**Code availability.** The codes that support the findings of this study are available from the corresponding author on reasonable request.

## Methods

**Materials.** The single-crystalline 1T-TaS$_2$ samples were grown by the chemical vapor transport method from elemental Ta and S with iodine as a transport agent[27]. The bulk crystals were exfoliated by the scotch tape technique and immediately inserted into the vacuum chamber to avoid oxidation. The X-ray diffraction confirmed the high quality of the samples. DC transport measurements were performed for the initial characterization of the samples.

**EUV transient grating experimental setup.** The experimental setup is shown schematically in Fig. 1a. The TG experiments were performed at the EIS-TIMER beamline at the FERMI FEL facility (Triste, Italy) in backward diffraction geometry. ~50 fs EUV pulses from the FEL source are split by the wavefront division beam splitters into two pump and one probe beams. After the spectral filtering, the beams are overlapped and focused on the sample, which is mounted on a cold finger cryostat. The angle between the two pump beams was set to 27.6°, resulting in an interference pattern on the sample surface with a period of 87 nm, taking into account the pump wavelength $\lambda_{pump} = 39.9$ nm. The state of the sample is then monitored as a function of the time delay between the pumps and the probe in a stroboscopic manner, by recording the intensity of diffracted probe beam ($\lambda_{pr} = 13.3$ nm) with the CCD camera. For studying metastable structures at low temperatures the sample was excited by a train of ~50 pump pulses once (corresponding to 1 s illumination time), and then the slow dynamics were monitored in real time by repeatedly exposing the sample to the probe beam only. The incident fluence on the sample was 3.5 mJ/cm² for the probe beam and 3.8 mJ/cm² for each of the pump beams.

An energy level diagram showing the transitions for pumping and probing is illustrated in Fig. 1e. The pump photon energy $\hbar\omega_{pump} = 31.1$ eV has nearby resonances in 1T-TaS$_2$[41], with initial states of Ta whose binding energies are 34.7 (2) eV ($O_3$ edge $5p_{3/2}$), 25.5 (1.7) eV ($N_6$ edge $4f_{5/2}$) and 23.6 (1.2) eV ($N_7$ edge $4f_{7/2}$) respectively, and $5d_{xy}, 5d_{3z^2-1}$ Ta final state bands (with bandwidth ~2 eV) near the Fermi level. The full width at half maximum (in eV) are indicated in brackets. The probe wavelength $\hbar\omega_{pr} = 93.3$ eV is non-resonant and has at least 3 times longer absorption length, which ensures that it does not disturb the grating despite the relatively high incident fluence used in the experiment.

The experiments were performed at room temperature (in the NC state) and 100 K (in the C ground state).

**Fitting procedure.** The room temperature data was fitted as a superposition of exponential background and two damped oscillators:

$$I_{\text{TG}}(t) = I_0 \text{H}(t)\left(1 - e^{t/\tau_0}\right)\left[I_1 e^{t/\tau_1}\sin(\omega_1 t + \varphi_1) + I_2 e^{t/\tau_2}\sin(\omega_2 t + \varphi_2) + I_3 e^{t/\tau_3}\right]$$

where $I_n$ are amplitudes of individual components, $\omega_{1,2}$, $\varphi_{1,2}$, $\tau_{1,2}$ are the frequencies, phases and damping constants of the oscillators, and $\tau_3$ is the relaxation time of the exponential background. Additional exponent with the characteristic time $\tau_0$ was used to account for the rise time of the signal. Since the FEL pulse length is much shorter than the relevant time scales, we assumed instantaneous excitation, which was modeled by a Heaviside function $\text{H}(t)$.